\documentclass[preprint,12pt]{elsarticle}
\usepackage{amssymb}
\usepackage{amsmath}
\usepackage{url}
\usepackage{graphicx}

\def\be{\begin{equation}}
\def\ee{\end{equation}}
\def\bea{\begin{eqnarray}}
\def\eea{\end{eqnarray}}
\def\bdm{\begin{displaymath}}
\def\edm{\end{displaymath}}
\journal{Nuclear Physics A}

\begin{document}

\begin{frontmatter}

\title{Investigation of Gamow Teller Transition Properties in $^{56-64}$Ni Isotopes Using QRPA Methods}

\author{Sadiye CAKMAK$^{1}$}
\author {Jameel-Un Nabi$^{2}$}
\author {Tahsin BABACAN$^{3}$}
\address{$^1$Program of Medical Monitoring
Techniques, Osmangazi University,Eskisehir,Turkey}

\address{$^2$Faculty of Engineering Sciences, GIK Institute of Engineering
Sciences and Technology, Topi 23640, Swabi, Khyber Pakhtunkhwa,
Pakistan}
\address{$^3$Department of Physics, Celal Bayar University, Manisa, Turkey}


\begin{abstract}
Weak rates in nickel isotopes play an integral role in the dynamics
of supernovae. Electron capture and $\beta$-decay of nickel
isotopes, dictated by Gamow-Teller transitions, significantly alter
the lepton fraction of the stellar matter. In this paper we
calculate Gamow-Teller (GT) transitions for isotopes of nickel,
$^{56-64}$Ni, using QRPA methods. The GT strength distributions were
calculated using four different QRPA models. Our results are also
compared with previous theoretical calculations and measured
strength distributions wherever available. Our investigation
concluded that amongst all RPA models, the pn-QRPA(C) model best
described the measured GT distributions (including total GT strength
and centroid placement). It is hoped that the current investigation
of GT properties would prove handy and may lead to a better
understanding of the presupernova evolution of massive stars.

\end{abstract}


\end{frontmatter}

\section{Introduction}
\label{intro} Weak interaction reactions, including electron capture
and $\beta$-decay, have a significant role in finding out the
working mechanism of key astrophysical events. These include, but
are not limited to, presupernova evolution of massive stars,
dynamics of supernovae, nucleosynthesis problem   (e.g.
\cite{Arn96,Sarr03}). Electron capture and $\beta$-decay directly
affect the lepton to baryon ratio, Y$_{e}$, and entropy at
presupernova stages. Charge-changing transitions usually termed as
Gamow-Teller (GT) transitions convert a proton (neutron) into a
neutron (proton). Weak rates of iron-regime nuclei are mostly
required to understand these key astrophysical processes (see for
example \cite{Ful80,Auf94,Nab99,Lan00,Nab04,Nab13}). Studies of
presupernova evolution of massive stars are available where electron
capture and $\beta$-decay rates of nickel isotopes in stellar matter
are thought to lead these stars to supernova explosion (e.g.
\cite{Auf94,Heg01}). The $\beta$-decay and electron capture of
nuclei along the $s$- and $r$- processes determine the paths and
abundances of the elements synthesized in nucleosynthesis.

Early half-life studies of nickel isotopes were performed in 1990
\cite{Ber90} and 1998 \cite{Am98}. In  these papers, the authors
determined the half life of nickel isotopes with mass number $A>69$.
Franchoo et al. \cite{Fran98} used laser ions enabled sources, for
the first time, and performed sensitive measurements of $\beta$ and
$\gamma$ decays of $^{68-74}$Ni isotopes. In 2000, a study by
Brachwitz et al. \cite{Brach00} claimed that the complete duration
of electron capture on Ni isotopes, in the area near $^{56}$Ni,
assisted in a better understanding of the thermo-nuclear supernovae
(type 1a).  Heger and collaborators \cite{Heg01} investigated the
presupernova evolution of massive stars and found out that electron
capture rates on Ni isotopes have a major part to play in time
evolution of electron to baryon ratio in the 25-40 solar mass star
explosions. Sasano et al. \cite{Sas11} used $^{56}$Ni (p,n)
reactions to determine the intensity of GT for the Ni isotope in the
direction $\Delta T_{z}=-1$. In Fermi transitions, a parent state is
connected to a single daughter state (isobaric analog state) whereas
the total GT strength is fragmented and connected to many daughter
states. This feature makes both the measurement and calculation of
GT strength a challenging task.

There has been various experimental studies of GT strength
distributions via (n,p)\cite{Kat94}, (p,n)\cite{Sas11}, (d,$^2$He)
\cite{Col06}, (t,$^3$He) \cite{Col06} and ($^3$He,t)
\cite{Fuj02,Fuj07} charge-exchange reactions. The findings hint
toward the fact that the measured total GT strength is quenched
(when compared with theoretical predictions) and fragmented over
many final states in daughter nucleus. The data also shows a
misplacement in the GT centroid used in the parameterizations of
\cite{Ful80}.  On the theoretical front, shell model/RPA
calculations (using various interactions e.g.  KB3G \cite{Pov01},
FPD6 \cite{Ric91}, GXPF1 \cite{Suz09}, GXPF1a \cite{Sarr03}), large
scale shell model calculation (with KB3G \cite{Col06,Cau99}, GXPF1
\cite{Col06} interactions), pn-QRPA (proton-neutron quasi random
phase approximation) calculations \cite{Sarr03,Nab99,Nab04,Nab99a},
HF+BCS (Hartree-Fock+Bardeen-Cooper-Schrieffer) and HF+BCS+QRPA
\cite{Sarr03} are the key calculations that are also microscopic in
nature. In a recent study by Rahman and Nabi \cite{Rah14}
proton-rich nuclei in mass range $46\leq A \leq 56$ were focused and
the GT strength distribution function and the electron capture rates
were calculated.

There exists a fair amount of measured data and theoretical studies
on GT transition of Ni isotopes in the mass region of 56--64 in
literature. We focused on the investigation of GT transition
properties for Ni isotopes in this mass region. In this work we
calculate GT strength distributions for isotopes of nickel
(56$\leq$A$\leq$64) by using different QRPA formalisms namely
pn-QRPA model \cite{Hal67,Kru84}, Pyatov Method (PM) and Schematic
Model (SM) \cite{Pya77,Bab05,Bab05a}. Schematic model is a special
case of PM and will be discussed later in the next section. These
models were explained and further classified into eight sub-models
in \cite{Cak14}. Findings of \cite{Cak14} were that PM(B) (separable
GT forces in both particle-particle and particle-hole channels in
\textit{spherical} basis), SM(B) (separable GT forces in both
particle-particle and particle-hole channels in \textit{spherical}
basis), SM(C) (separable GT forces in both particle-particle and
particle-hole channels in \textit{deformed} basis) and
pn-QRPA(C)(separable GT forces in both particle-particle and
particle-hole channels in \textit{deformed} basis performed in a
multi-shell single-particle space) were the four better models out
of the eight studied for a better description of the GT strength
functions. Here we use these four QRPA approaches (namely PM (B), SM
(B), SM (C) and pn-QRPA (C)) to calculate GT strength distributions
of nickel isotopes in mass range 56-64.

Brief and necessary theoretical formalism is presented in the next
section. During the early phases of presupernova evolution, the
electron chemical potential is of the same order of magnitude as the
nuclear $Q$ value and the weak rates are sensitive to the details of
the low-lying GT$_{\pm}$ strength functions in daughter nuclei. With
proceeding collapse the stellar density increases by orders of
magnitude and the resulting electron chemical potential is
significantly higher than the nuclear $Q$ value and then the stellar
rates are largely determined by the centroid and the total strength
of the GT$\pm$ strength functions. As such reliable theoretical estimate
of \textit{both} low-lying GT distributions (at low stellar
temperatures and densities) and total GT strength/centroid (at
relatively high temperatures and densities) are required for an
accurate calculation of stellar weak rates. In the third section, we
compare  our GT$_{\pm}$ strength distributions results of nickel
isotopes with previous theoretical and measured results. We also
calculate and compare centroid (and width) of  calculated GT
distributions in this section. In the last section, we summarize our
investigation and present important conclusions.

\section{Theoretical Formalism}
\label{sec:1} In this section we present a brief summary of the
necessary formalism for the three models used in this work. For
further details we refer to \cite{Cak14}.
\subsection{The Schematic Model (SM) and the Pyatov Method (PM)}

The Hamiltonian of schematic model (SM) is given by
\begin{eqnarray}
H_{SM}=H_{SQP}+h_{ph}+h_{pp}, \label{Eqt. 1}
\end{eqnarray}
where, $H_{SQP}$ is the single quasiparticle Hamiltonian,
$h_{ph}^{GT}$ and $h_{pp}^{GT}$ are the GT effective interactions in
the particle-hole and particle-particle channels, respectively
\cite{Cak14,Nec10}. The effective interaction constants in the two
channels were fixed from experimental value of the GT resonance
energy. Terms not commuting with GT operator were removed from the
total Hamiltonian. The mean field approximation was renewed by
adding an effective interaction term $h_{0}$ \cite{Cak14} given as
\begin{eqnarray}
h_{0}=\sum_{\rho=\pm}\frac{1}{2\gamma_{\rho}}\sum_{\mu=0,\pm1}[H_{sqp}-V_{c}-V_{ls}-V_{1},G_{1\mu}^{\rho}]^{\dagger}\cdot\nonumber
\end{eqnarray}
\begin{eqnarray}
[H_{sqp}-V_{c}-V_{ls}-V_{1},G_{1\mu}^{\rho}] \label{Eqt. 2}.
\end{eqnarray}
The strength parameter $\gamma_{\rho}$ of the effective interaction
was found from necessitating the commutation conditions (for details
see \cite{Cak14,Sal06})
\begin{eqnarray}
\gamma_{\rho}=\frac{\rho}{2}\langle0|[[H_{sqp}-V_{c}-V_{ls}-V_{1},G_{1\mu}^{\rho}],G_{1\mu}^{\rho}]
|0\rangle.\nonumber
\end{eqnarray}
The GT operator $G_{1\mu}^{\pm}$, which commutes with the
Hamiltonian, is given as
\begin{eqnarray}
G_{1\mu}^{\pm}=\frac{1}{2}\sum_{k=1}^{A}[\sigma_{1\mu}(k)t_+(k)+\rho(-1)^\mu\sigma_{1-\mu}(k)t_-(k)]
      (\rho=\pm1),  \label{Eqt. 3}
\end{eqnarray}
where $\sigma_{1\mu}(k)=2s_{1\mu}(k)$ are the spherical components
of the Pauli operators, $t_\pm=t_x(k)\pm it_y(k)$ are the isospin
raising/lowering operators.

The total Hamiltonian of Pyatov Method is
\begin{eqnarray}
H_{PM}=H_{SQP}+h_{0}+h_{ph}+h_{pp} \label{Eqt. 4}.
\end{eqnarray}
The GT transition strengths were calculated by summing the nuclear
matrix elements

\textbf{\begin{eqnarray}
B_{GT}^{(\pm)}(\omega_{i})=\sum_{\mu}|M_{\beta^{\pm}}^{i}(0^{+}\longrightarrow1^{+}_{i}|^{2},
\label{Eqt. 5}
\end{eqnarray}}
where $\omega_{i}$ are the excitation energies in the daughter
nucleus.
 The $\beta^{\pm}$ transition strengths were finally
calculated using
\begin{eqnarray}
B(GT)_{\pm}=\sum_{i}B_{GT}^{(\pm)}(\omega_{i}). \label{Eqt. 6}
\end{eqnarray}
The calculated GT strengths should fulfill the Ikeda sum rule (ISR)
\cite{Ike64}
\begin{eqnarray}
ISR=B(GT)_{-}-B(GT)_{+}\cong3(N-Z). \label{Eqt. 7}
\end{eqnarray}
In our PM(B) and SM(B) models, the mean field Hamiltonians were
defined \textit{only} in particle space. Pairing interaction
potential (in both $pp$ and $ph$ channels) was added as a
perturbation term. However in our SM(C) model, mean field
Hamiltonian was identified in quasiparticle space by introducing
special Bogoliubov transformation.  In PM(B), symmetry
deteriorations, because of mean field approach, were restored by the
introduction of efficient interaction term $h_{0}$, In SM(B) and
SM(C) models, beta decay interactions were added without restoring
symmetry deteriorations. Calculated GT strength values from PM and
SM can have differences because of the effective interaction term
($h_{0}$) (for further details, see
\cite{Nec10,Bab05,Bab05a,Sal03,sel04}).

\subsection{The  proton-neutron quasi particle
random phase approximation (pn-QRPA) Model}

For the pn-QRPA model,the Hamiltonian  is given by
\begin{equation}
H^{QRPA} = H^{sp} + V^{pair} + V ^{ph}_{GT} + V^{pp}_{GT}.
\label{Eqt. 8}
\end{equation}
Wave functions and single particle energies were calculated in
deformed Nilsson basis. BCS approximation was used to take care of
the pairing interaction in nuclei. The GT strength distribution
calculation was performed in a multi-shell single-particle space
with a schematic interaction \cite{Mut92}.

In the pn-QRPA formalism, GT transitions are expressed in terms of
phonon creation operators defined as
\begin{equation}
A^{+}_{\omega}(\mu) =
\sum_{pn}(X^{pn}_{\omega}(\mu)a^{+}_{p}a^{+}_{\overline{n}} -
Y^{pn}_{\omega}(\mu)a_{n}a_{\overline{p}}). \label{Eqt. 9}
\end{equation}
The sum in Eq.~(\ref{Eqt. 9}) runs over all proton-neutron pairs
with $\mu = m_{p} - m_{n}= -1, 0, 1,$ where $m_{p/n}$ denotes the
third component of the angular momentum.

The proton-neutron residual interaction occurs through $pp$ and $ph$
channels. Both the interaction terms can be given a separable form.
Both the $ph$ force given by
\begin{equation}
V^{ph}_{GT} = 2\chi\sum_{\mu}(-1)^{\mu}Y_{\mu}Y^{+}_{-\mu},\nonumber
\end{equation}
with
\begin{equation}
Y_{\mu} = \sum_{j_{n}j_{p}} \langle
j_{p}m_{p}|t_{-}\sigma_{\mu}|j_{n}m_{n} \rangle
c^{+}_{j_{p}}m_{p}c_{j_{n}m_{n}}, \label{Eqt. 26}
\end{equation}
and the $pp$ interaction given by the separable force
\begin{equation}
V^{pp}_{GT} =
2\kappa\sum_{\mu}(-1)^{\mu}P_{\mu}P^{+}_{-\mu},\nonumber
\end{equation}
 with
\begin{equation}
P_{\mu}^{+} = \sum_{j_{n}j_{p}} \langle
j_{n}m_{n}|(t_{-}\sigma_{\mu})^{+}|j_{p}m_{p} \rangle \nonumber
\end{equation}
\begin{equation}
\times
(-1)^{l_{n}+j_{n}-m_{n}}C^{+}_{j_{p}m_{p}}C^{+}_{j_{n}-m_{n}},
\label{Eqt. 27}
\end{equation}
are taken into account in pn-QRPA(C) and SM(C) models.

The reduced GT transition probabilities in the pn-QRPA(C) model was
finally given by
\begin{equation}
B_{GT}^{\pm}(\omega) = |\langle \omega, \mu
\|t_{\pm}\sigma_{\mu}\|QRPA\rangle|^{2}, \label{Eqt. 10}
\end{equation}
where the symbols have their usual meanings and daughter excitation
energy is denoted by $\omega$. Eq.~(\ref{Eqt. 10}) gives the
measured values of $\beta^{\pm}$ transition strengths and also
satisfy the ISR (Eq.~(\ref{Eqt. 7})). For details of the pn-QRPA
model see \cite{Mut92,Hir93}.

The pn-QRPA(C) calculation was performed in a deformed basis.
Measured values of the deformation parameter ($\beta_2$) for
even-even isotopes of nickel, determined by relating the measured
energy of the first $2^{+}$ excited state with the quadrupole
deformation, were taken from \cite{Ram87}. For remaining cases the
deformation of the nucleus was calculated using
\begin{equation}
\beta = \frac{125(Q_{2})}{1.44 (Z) (A)^{2/3}},
\end{equation}
where $Z$ and $A$ are the atomic and mass numbers, respectively.
$Q_{2}$ is the electric quadrupole moment and the values were taken
from \cite{Moe81}. The Q-values of the charge-changing reactions
were taken from the recent mass compilation \cite{Aud12,Aud12i}.

\section{GT$_{\pm}$ Strength Distributions}

\label{sec:2} The GT transitions play an integral role in the
presupernova evolution of massive stars, during core-collapse and
finally when the stars go supernova. Consequently GT strength
distributions in nickel isotopes can be very important from
astrophysical viewpoint. At high stellar densities, of the order of
10$^{11}$ gcm$^{-3}$, the nuclear Q value and electron chemical
potential are roughly same order of magnitude. Under such physical
conditions the $\beta$-decay rates are sensitive to the details of
GT strength distributions. As the stellar cores stiffens to order of
magnitude larger densities, the chemical potential becomes much
larger than the Q-values and then the stellar weak rates may well be
dictated by total GT strength. Hence a microscopic calculation of GT
strength distributions is a prerequisite to understand the working
mechanism of these complex astrophysical phenomena.

In this section, we calculate and investigate the GT transition
properties of $^{56-64}$Ni isotopes. We present our calculations for
both GT strength distributions and total strength. Further we also
compare with previous calculations and measurements wherever
possible. We perform our calculations using the four QRPA models,
namely SM (B), PM (B), SM (C) and pn-QRPA (C) models, as discussed
in the Introduction.

\begin{figure}
\begin{center}
\resizebox{0.8\textwidth}{!}{
\includegraphics{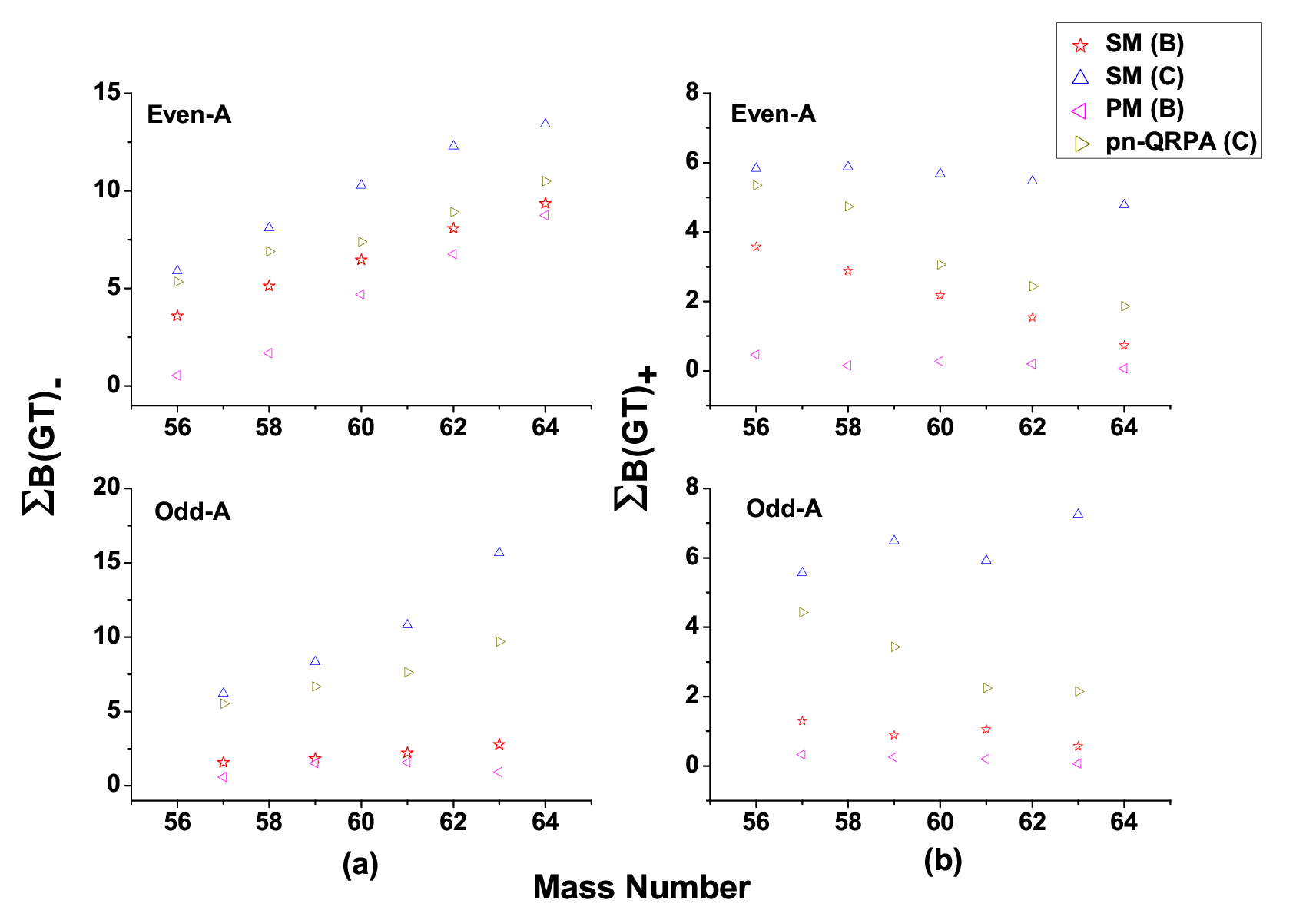}}
\end{center}
\caption{Total GT strength calculated in all four QRPA models. Even
mass nuclei are shown in the upper panel. Lower panels depict the
calculated strength for odd mass nuclei.} \label{fig:1}
\end{figure}

\begin{figure}
\begin{center}
\resizebox{0.8\textwidth}{!}{
\includegraphics{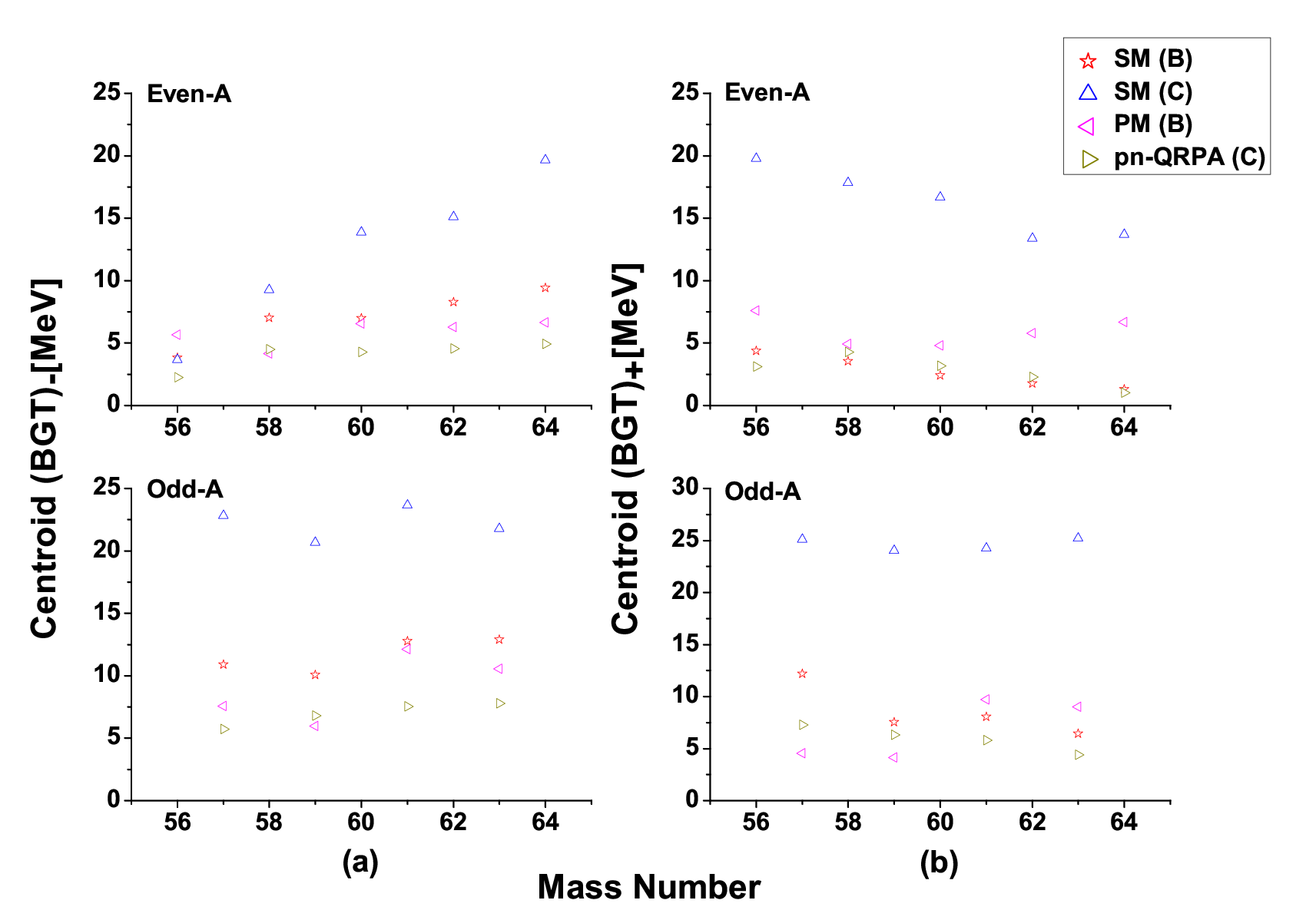}}
\end{center}
\caption{Comparison of centroid of calculated GT distributions,
along $\beta$-decay and electron capture directions, in all four
QRPA models.} \label{fig:2}
\end{figure}

\begin{figure}
\begin{center}
\resizebox{0.8\textwidth}{!}{
\includegraphics{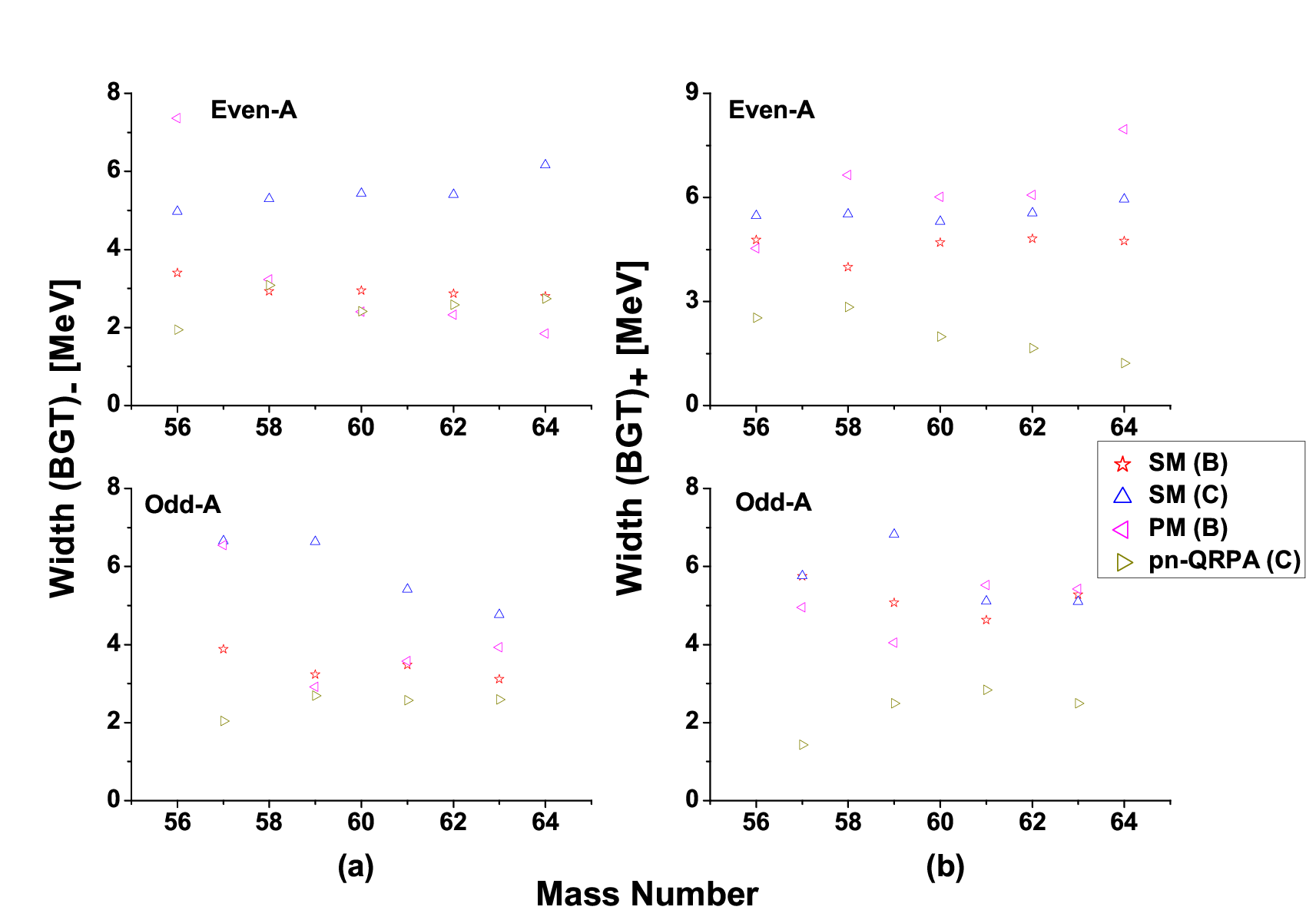}}
\end{center}
\caption{Comparison of width of calculated GT distributions, along
$\beta$-decay and electron capture directions, in all four QRPA
models.} \label{fig:3}
\end{figure}

We present our calculated total GT strength distributions along with
calculated centroids and widths in Figs.~\ref{fig:1}~-~\ref{fig:3}.
Fig.~\ref{fig:1} displays the calculated total GT strength  for the
four chosen QRPA models. The left panels show calculated strength
along $\beta$-decay direction whereas the left panels show the same
along electron capture direction. In order to highlight the effect
of deformation we have divided our calculation into even mass (where
the deformation parameter approaches zero) and odd mass isotopes of
nickel as shown in Fig.~\ref{fig:1}. The spherical calculations lead
to smaller strength functions. It can be seen that the highest
strength is calculated by SM (C) model and lowest by PM (B).  It is
also to be noted that for odd mass nuclei the deformed calculations
are significantly bigger than the spherical calculations (specially
along $\beta$-decay direction). Our calculation indicates that
incorporation of deformation into QRPA calculation lead to bigger GT
strength values.

We calculated centroids and widths of calculated GT strength
distributions (discrete in nature) in all our four QRPA models.
Mathematically the centroid of the calculated GT strength
distribution was calculated as
\begin{equation} Centroid(BGT)_{\pm} =
\frac{\sum_{j} E_{j}\times B_{j}(GT_{\pm})}{\sum_{j}
B_{j}(GT)_{\pm}}.
\end{equation}
where $E_{j}$ are the daughter excitation energies in units of MeV
and $B_{j}(GT)_{\pm}$ are the  calculated GT strength along electron
capture and $\beta^{-}$ directions, respectively (in arbitrary
units).

The width of ($GT_{\pm}$) strength distribution function was
calculated using the formula
\begin{equation} Width(BGT)_{\pm} = \sqrt{\frac{\sum
(E_{j}-\overline{E}_{\pm})^{2}\times B_{j}(GT_{\pm})}{\sum
B_{j}(GT)_{\pm}}} .
\end{equation}
where the $\overline{E}_{\pm}$ are the centroids calculated as
discussed above.

Comparison of centroids is shown in Fig.~\ref{fig:2}. Here one notes
that whereas both SM(C) and pn-QRPA(C) models calculated bigger GT
strength, it is only the pn-QRPA(C) model in which the centroid of
GT distributions resides at low excitation energy in daughter
nucleus. This may transform to bigger weak-interaction rates once
the phase space functions are incorporated in the calculations.
SM(C) model in fact calculates biggest value of centroids of all the
four models. The pn-QRPA(C) model also calculates smallest widths
for the calculated GT strength distributions (Fig.~\ref{fig:3}).

The deviation of our calculated (re-normalized) Ikeda sum rule
values from theoretically value is presented for both even and odd
isotopes. In all our calculations, a quenching factor  of
$f_{q}^{2}$ = (0.6)$^{2}$ was used (as used normally in $fp$-shell
calculations). The (re-normalized) Ikeda sum rule in all four models
then transforms as
\begin{equation}
Re-ISR = B(GT)_{-}-B(GT)_{+}\cong 3f_{q}^{2}(N-Z). \label{Eqt. ISR}
\end{equation}

\begin{figure}
\begin{center}
\resizebox{0.8\textwidth}{!}{
\includegraphics{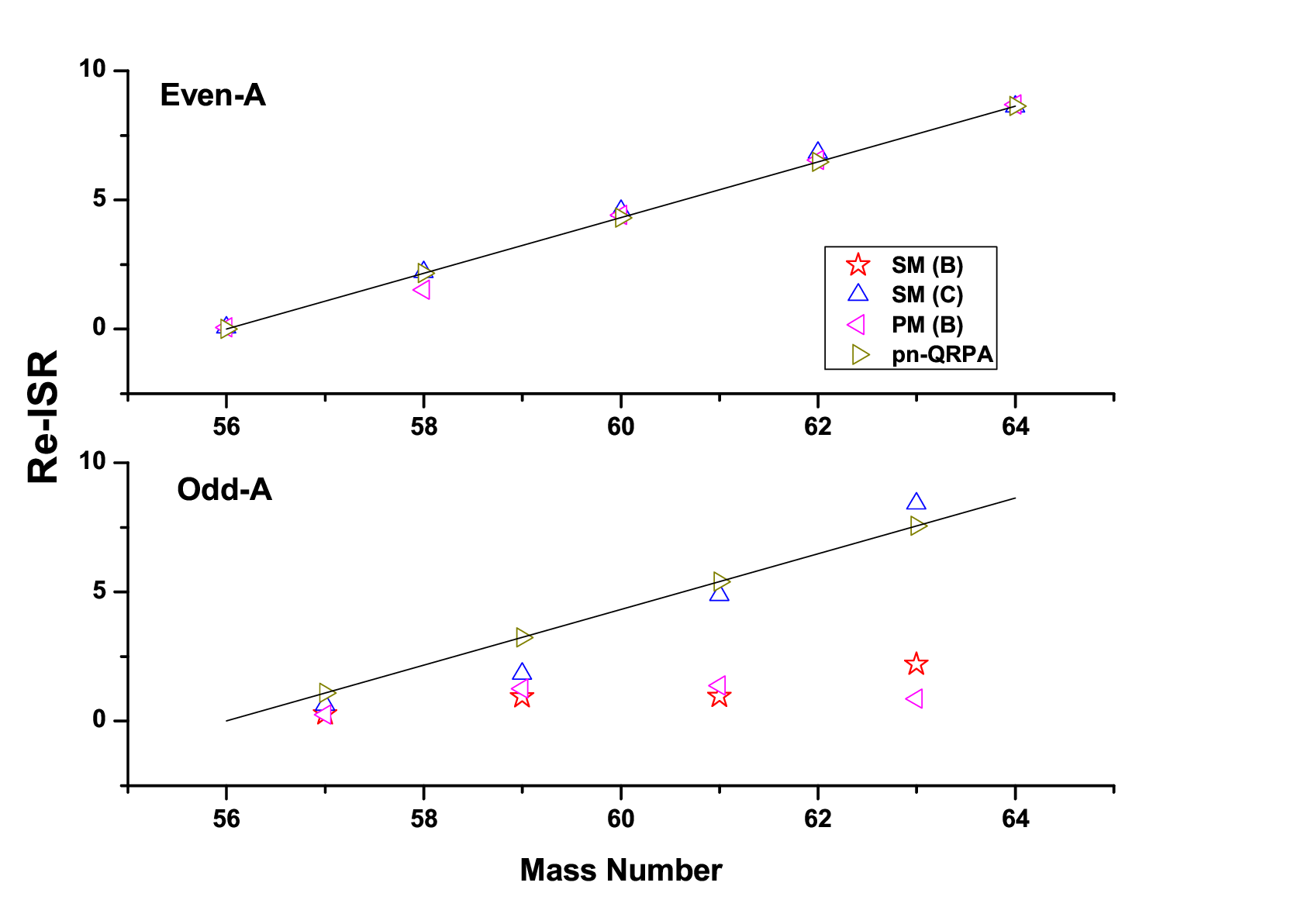}}
\end{center}
\caption{ Comparison of calculated re-normalized Ikeda sum rule in
all four QRPA models. Theoretically allowed sum rule is shown as a
straight line to guide the eye.} \label{fig:4}
\end{figure}

In Fig.~\ref{fig:4} the solid line shows the theoretical Ikeda sum
rule value. If we treat nucleons as point particles and ignore
two-body currents, the model independent Ikeda sum rule should be
satisfied by all calculations. For even-A isotopes we get a decent
agreement whereas for odd-A isotopes the spherical models fail. The
pn-QRPA(C) model satisfies well the sum rule. The remaining three
QRPA models struggle hard to satisfy the rule for odd-A cases. The
deviations increase with increasing mass number for SM(B), PM(B) and
SM(C) models. The $2p - 2h$ configuration mixing can account for the
missing strength in the PM and SM models (for further discussion see
\cite{Cak14,Cak15}).

\begin{figure}
\begin{center}
\resizebox{1.3\textwidth}{!}{
\includegraphics{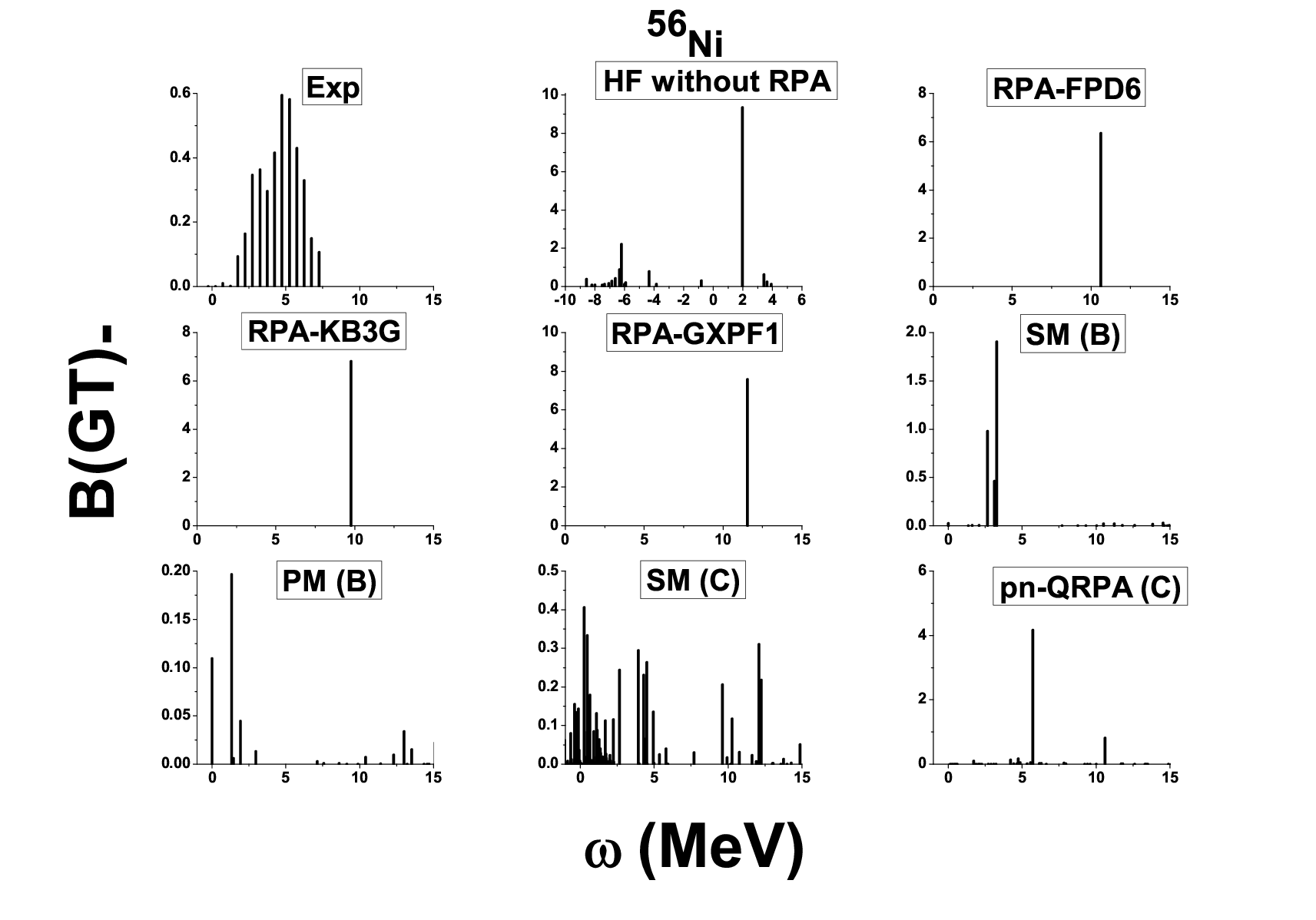}}
\end{center}
\caption{GT strength distributions in $^{56}$Ni along $\beta$-decay
direction. Exp shows measured distribution by \cite{Sas12}. Second
panel on top shows the calculated unperturbed GT strength without
RPA correlations by \cite{Bai13}. RPA calculations using shell model
interactions -- FDP6, KB3G and GXPF1 were taken from \cite{Nab13}.
Last four panels depict calculated strength distributions using our
four QRPA models. The abscissa represents energy in the daughter
nucleus. } \label{fig:5}
\end{figure}

\begin{figure}
\begin{center}
\resizebox{0.8\textwidth}{!}{
\includegraphics{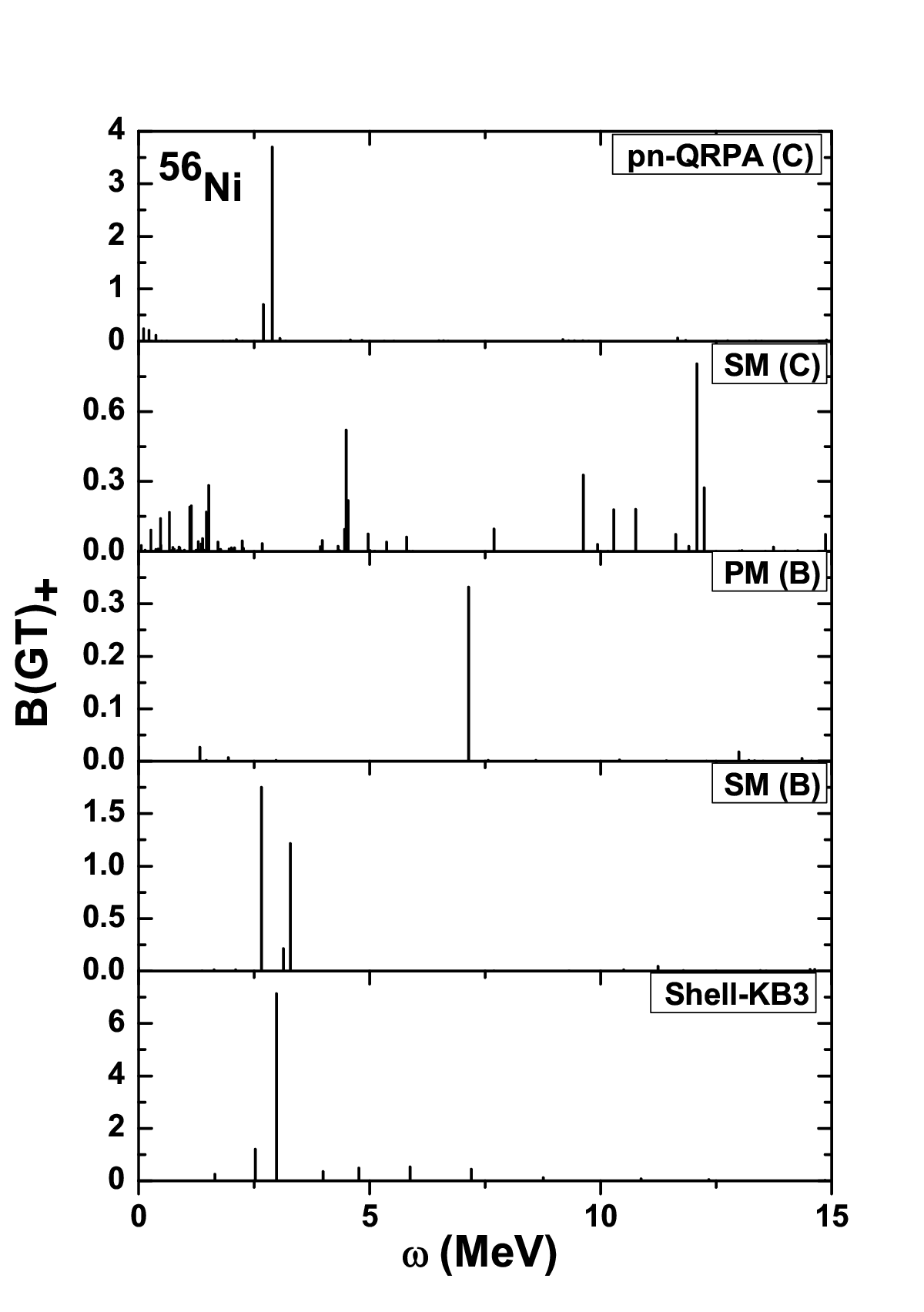}}
\end{center}
\caption{Calculated B(GT)$_{+}$ strength distributions in $^{56}$Ni
compared with shell model calculation \cite{Lan98}. The abscissa
represents energy in the daughter nucleus. } \label{fig:6}
\end{figure}

After presenting a mutual comparison of the four QRPA models we next
discuss how our calculated GT distributions compare with the
previous calculations and measurements. Fig.~\ref{fig:5} shows the
B(GT)$_{-}$ strength distributions in $^{56}$Ni isotopes. On the
experimental side we show the results of $^{56}$Ni$(p,n)$
charge-exchange reaction \cite{Sas12} labeled Exp.  We show the
Hartree-Fock (HF)  (unperturbed strength calculation) \cite{Bai13}
and RPA calculation using shell model interactions- FDP6
\cite{Ric91},  KB3G \cite{Pov01} and GXPF1 \cite{Hon02} performed by
\cite{Nab13} in the next four panels of Fig.~\ref{fig:5}. The
experimental data is not reproduced well by theoretical calculations
for this doubly magic nucleus. Whereas RPA calculations  put all
strength in one transition, the other models do show a fragmented
distribution. The centroids for the GT distribution calculated by SM
(B), PM (B), SM (C) and pn-QRPA (C) models are 3.86 MeV, 5.65 MeV,
3.66 MeV and 2.26 MeV, respectively. These values are close to the
experimental value of 4.1 MeV \cite{Sas12}. The RPA model
 calculates centroid values of 10.62 MeV, 9.77 MeV and
11.54 MeV with the FPD6, KB3G and GXPF1 interactions, respectively
\cite{Nab13}.  Calculated GT strength along the electron capture
direction for $^{56}$Ni is shown in Fig.~\ref{fig:6}.  Here we
compare our calculated GT distributions against the shell model
(with KB3 interaction) calculation \cite{Lan98}. We note that shell
model result is in very good agreement with the pn-QRPA (C) model
calculation. SM (C) strength is well fragmented. Suzuki and
collaborators \cite{Suz11} performed a shell model calculation of GT
strength distributions and stellar electron capture rates of nickel
isotopes. They used the KB3G and GXPF1J \cite{Hon05} interactions
for their calculation. It was argued in their calculation that the
GXPF1J interaction best reproduced the observed measured GT
strength, specially for neutron-rich nickel isotopes. The cumulative
GT strength distribution using the KB3G and GXPF1J interactions may
be seen from Fig.~5 of \cite{Suz11}. The total GT strength
calculated using the GXPF1J (KB3G) interaction was 6.2 (5.4). These
strength values compare well with the total GT strength value of
5.34 calculated using the pn-QRPA(C) model.

\begin{figure}
\begin{center}
\resizebox{1.3\textwidth}{!}{
\includegraphics{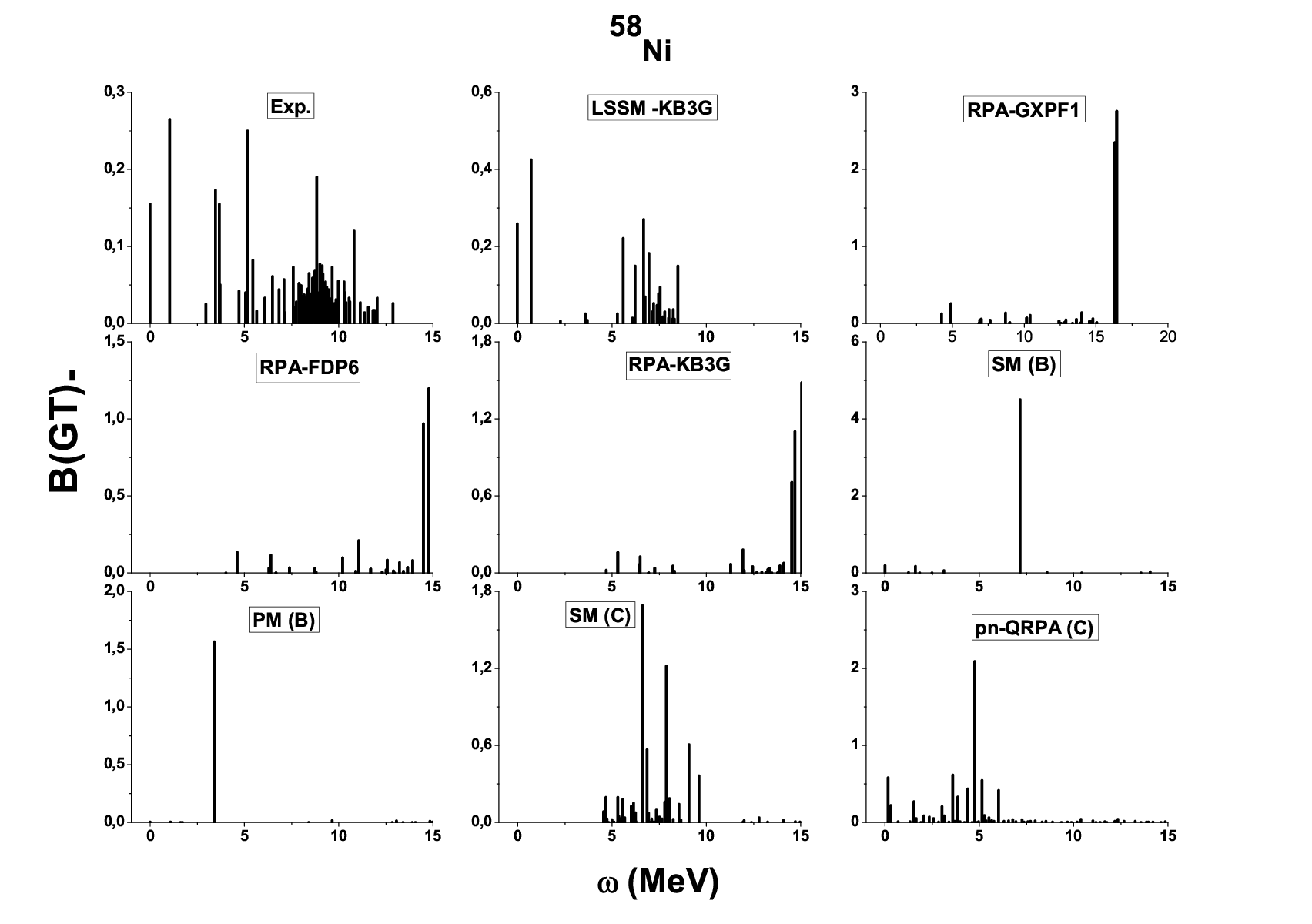}}
\end{center}
\caption{GT strength distributions in $^{58}$Ni along $\beta$-decay
direction. Exp. shows measured distribution by \cite{Fuj02,Fuj07}.
Second panel on top shows the large scale shell model calculation by
\cite{Cau99}. RPA calculations using shell model interactions --
GXPF1, FPD6 and KB3G  were taken from \cite{Nab13}. Last four panels
depict calculated strength distributions using our four QRPA models.
The abscissa represents energy in the daughter nucleus.}
\label{fig:7}
\end{figure}

The B(GT)$_{-}$ strength distributions in $^{58}$Ni is presented in
Fig.~\ref{fig:7}. The measured data was taken from the
$^{58}$Ni$(^{3}He,t)$ charge-exchange reaction \cite{Fuj02,Fuj07}.
Here we also show the results of Large Scale Shell Model calculation
(LSSM) with KB3G interaction \cite{Cau99} and RPA calculation with
KB3G, FDP6 and GXPF1 interactions \cite{Nab13}. The comparison shows
that the best agreement with experiment is seen in LSSM model. The
pn-QRPA(C) model also fairly describe the low-lying measured data.
The RPA calculation exhibits peaks at higher energies (around 14-16
MeV). In SM(B) and PM(B) models, only one main peak is seen.

\begin{figure}
\begin{center}
\resizebox{1.3\textwidth}{!}{
\includegraphics{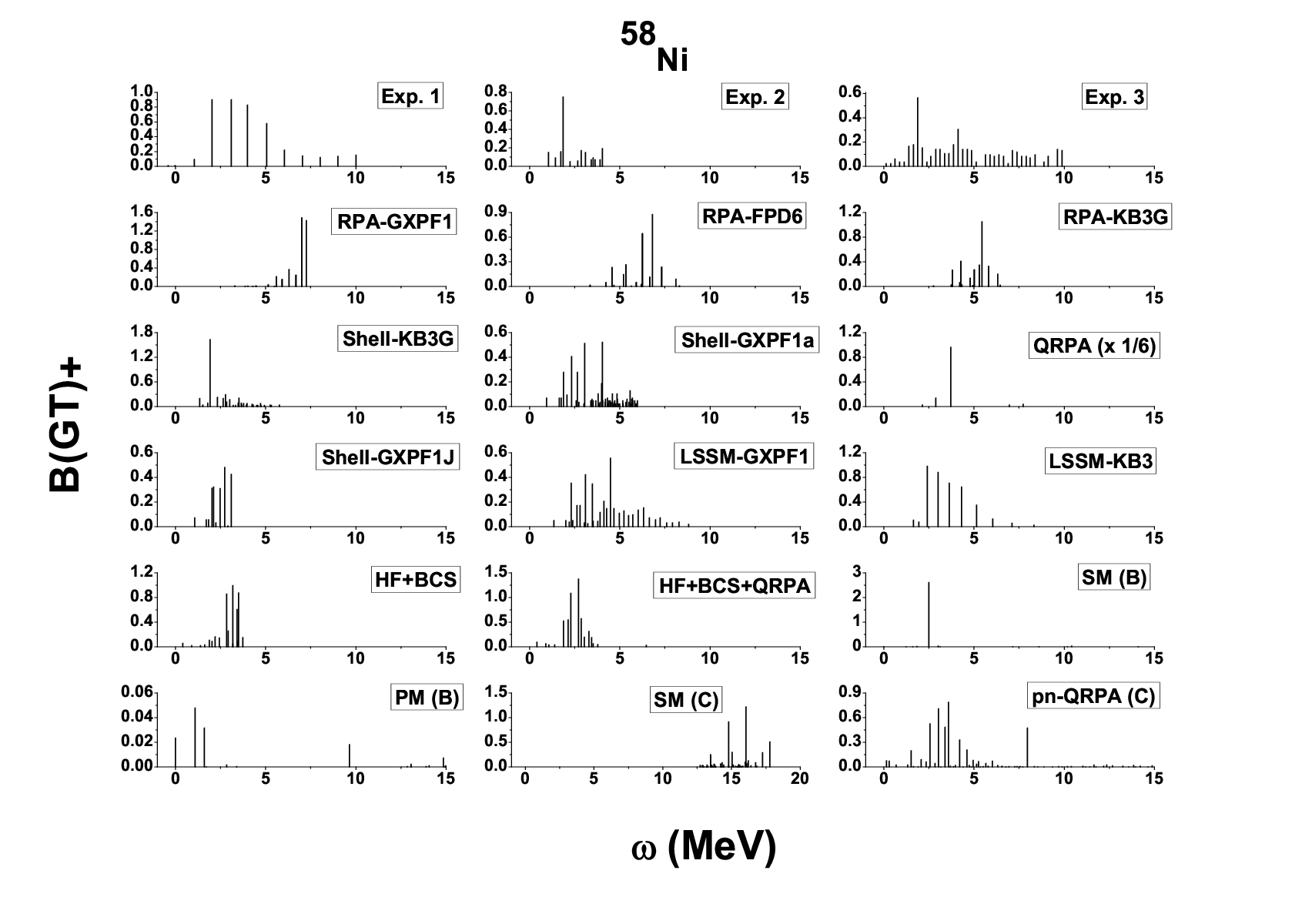}}
\end{center}
\caption{GT strength distributions in $^{58}$Ni along electron
capture direction. First row shows measured distributions by
\cite{Kat94,Hag04,Col06}, respectively. Second row depicts RPA
calculations using shell model interactions -- FDP6, KB3G and GXPF1
taken from \cite{Nab13}. Third row presents shell model (employing
KB3G and GXPF1a interactions) and QRPA calculations done by
\cite{Col12}. Fourth row shows shell model calculation using GXPF1J
interaction by \cite{Suz11} and large scale shell model calculation
using GXPF1 interaction by \cite{Col06} and KB3 interaction by
\cite{Cau99}. HF+BCS and HF+BCS+QRPA calculations were taken from
\cite{Sarr03}. Last four panels depict calculated strength
distributions using our four QRPA models. The abscissa represents
energy in the daughter nucleus.} \label{fig:8}
\end{figure}

We found a wealth of data for GT strength distribution of $^{58}$Ni
along the electron capture direction. We found three different
measurement results displaying rather varying strength
distributions. These are shown in top three panels of
Fig.~\ref{fig:8}. The $^{58}$Ni$(n,p)$ data at incident energy of
198 MeV is presented as Exp. 1 \cite{Kat94}. Exp. 2 shows the
$(d,^{2}He)$ charge-exchange reactions performed by Hagemann et al.
\cite{Hag04}. Exp. 3  depicts the $^{58}$Ni$(t,^{3}He)$
charge-exchange reaction performed by Cole and collaborators
\cite{Col06} whereas Besides we present theoretical calculations of
GT strength for $^{58}$Ni  using 15 different microscopic models.
For references see caption of Fig.~\ref{fig:8}. Here we note that
LSSM and pn-QRPA (C) are in reasonable agreement with the low-lying
measured data. This very good comparison with experimental data can
lead to very reliable estimate of stellar weak rates using the
pn-QRPA(C) model.

\begin{figure}
\begin{center}
\resizebox{1.25\textwidth}{!}{
\includegraphics{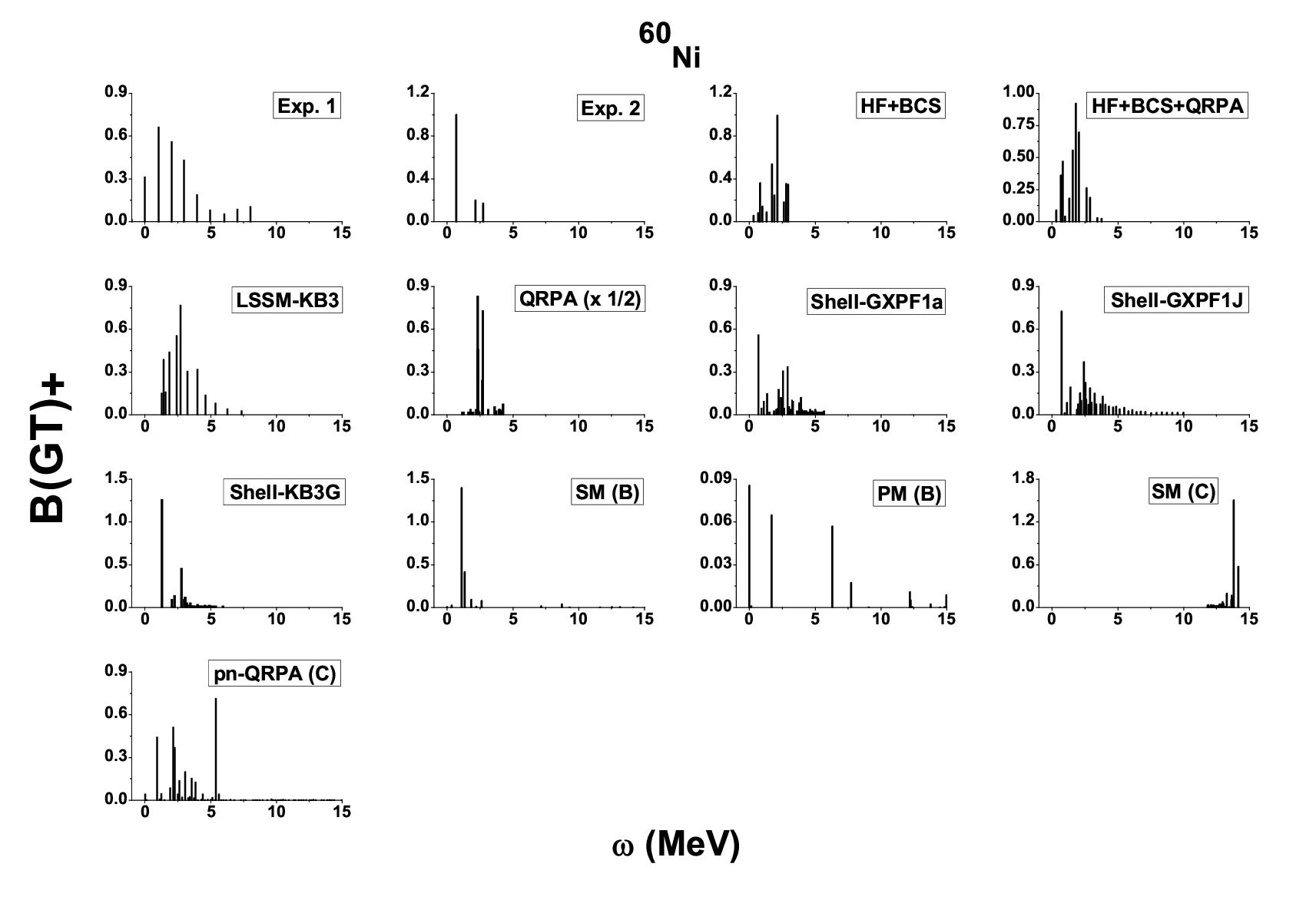}}
\end{center}
\caption{GT strength distributions in $^{60}$Ni along electron
capture direction. Exp. 1 shows measured distribution by
\cite{Will95} and Exp. 2 by \cite{An08}. HF+BCS and HF+BCS+QRPA
calculations were taken from \cite{Sarr03}. Large scale shell model
calculation using KB3 interaction was taken from \cite{Cau99}. QRPA
and shell model (employing KB3G and GXPF1a interactions)
calculations were taken from \cite{Col12} while those using GXPF1J
interaction was taken from \cite{Suz11}. Last four panels depict
calculated strength distributions using our four QRPA models. The
abscissa represents energy in the daughter nucleus.} \label{fig:9}
\end{figure}

The  B(GT)$_{+}$  distribution for $^{60}$Ni was measured using
$(n,p)$ reaction at 198 MeV by Williams and collaborators
\cite{Will95}. The result is shown as Exp. 1 in Fig.~\ref{fig:9}.
Later Anantaraman et al. \cite{An08} studied the electron capture
strength for $^{60,62}$Ni using $(p,n)$ reactions at 134.3 MeV in
inverse kinematics at a higher resolution. The authors reported
differences in the two measurement results. Their findings are
marked as Exp. 2 in Fig.~\ref{fig:9}. The next two panels show the
HF approximation and HF+QRPA calculation built on a deformed
selfconsistent mean field basis employing pairing correlations in
BCS approximation \cite{Sarr03}. The basis were obtained from
two-body density-dependent Skyrme forces. This is followed by the
large scale shell model calculation of \cite{Cau99}. The QRPA and
shell model calculations (performed in a truncated $pf$ model space
comprising of 5 holes in the $f_{7/2}$ shell using the GXPF1a and
KB3G interactions) were taken from \cite{Col12}. Shell model
calculation \cite{Suz11} using the GXPF1J interaction is also shown.
The last four panels show our QRPA results. It is noted that
LSSM-KB3 calculation is in decent agreement with Exp. 1 whereas SM
(B) is in good agreement with Exp. 2. The pn-QRPA(C) distribution is
in good agreement with Exp. 1 till 5 MeV.

\begin{figure}
\begin{center}
\resizebox{1.25\textwidth}{!}{
\includegraphics{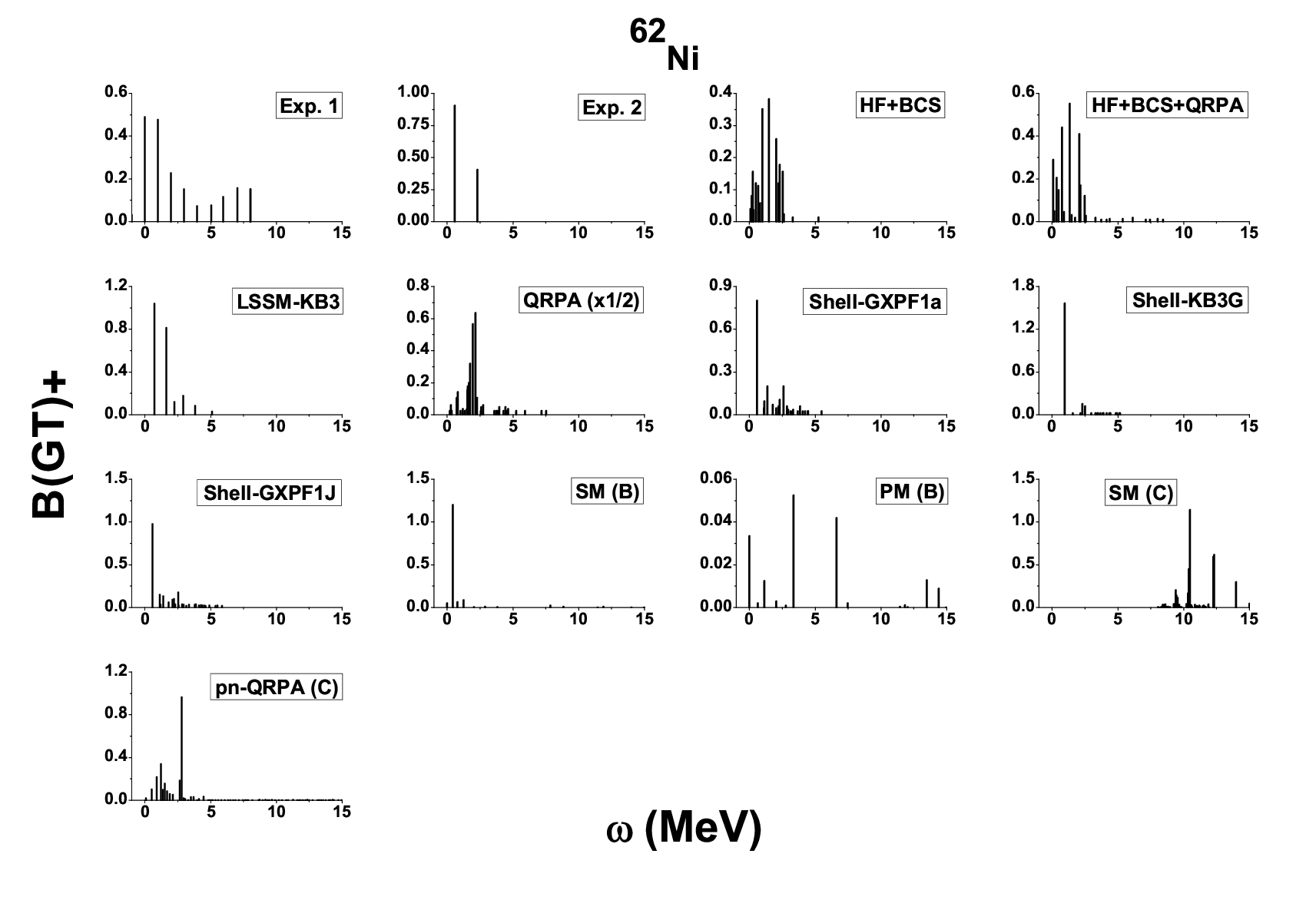}}
\end{center}
\caption{GT strength distributions in $^{62}$Ni along electron
capture direction. Exp. 1 shows measured distribution by
\cite{Will95} and Exp. 2 by \cite{An08}. HF+BCS and HF+BCS+QRPA
calculations were taken from \cite{Sarr03}. Large scale shell model
calculation using KB3 interaction was taken from \cite{Cau99}. QRPA
and shell model (employing KB3G and GXPF1a interactions)
calculations were taken from \cite{Col12}. Shell model calculation
using GXPF1J interaction was taken from \cite{Suz11}. Last four
panels depict calculated strength distributions using our four QRPA
models. The abscissa represents energy in the daughter nucleus.}
\label{fig:10}
\end{figure}

For $^{62}$Ni isotope, measured and calculated B(GT)$_{+}$
distributions were taken from same sources as in Fig.~\ref{fig:9}.
All the strength distributions are labeled same and presented in
Fig.~\ref{fig:10}.  The LSSM-KB3 data compares fairly good with Exp.
2 data. Among our four modes, it is the pn-QRPA (C) model which best
describes the experimental data.

\begin{figure}
\begin{center}
\resizebox{0.8\textwidth}{!}{
\includegraphics{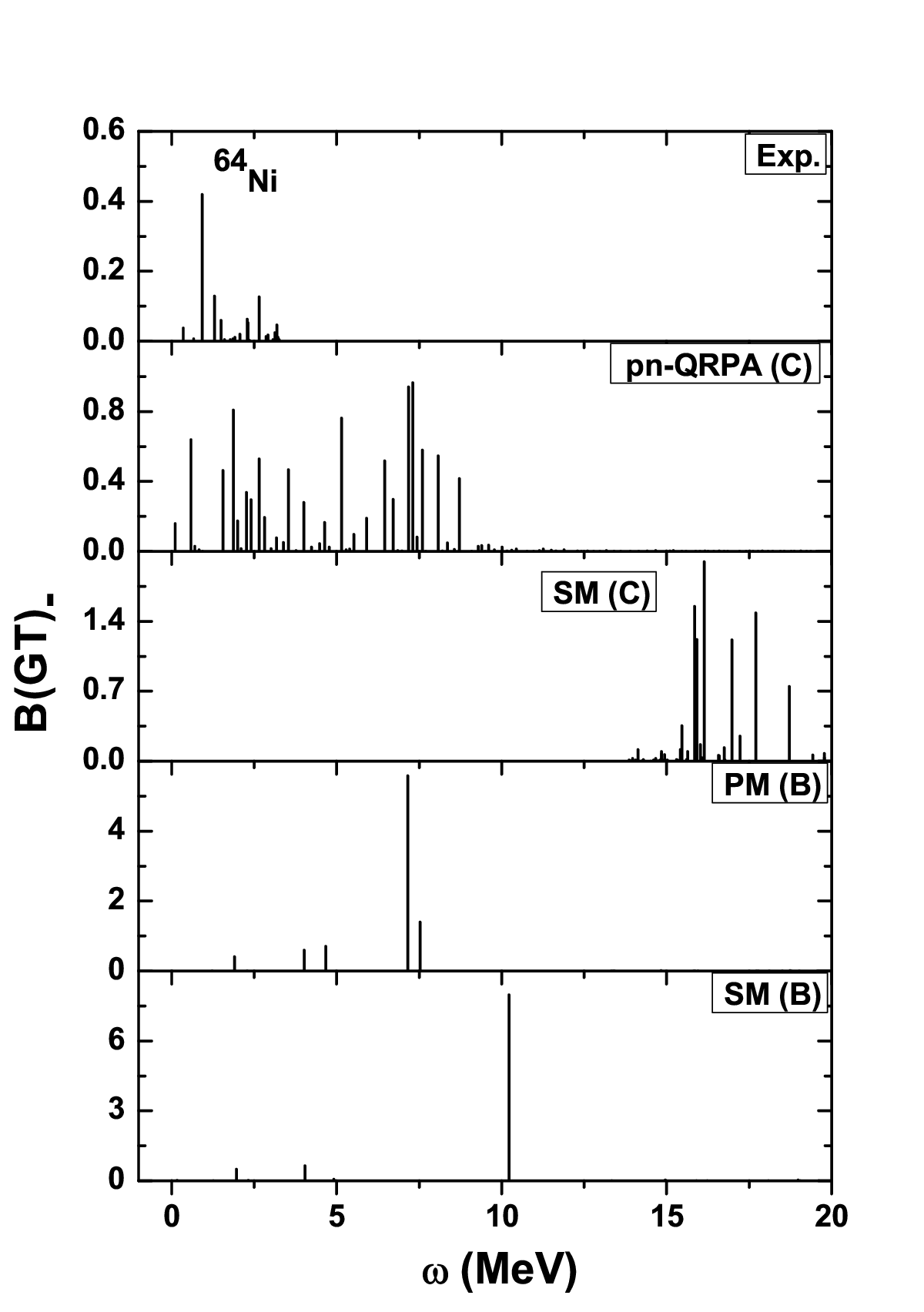}}
\end{center}
\caption{Calculated B(GT)$_{-}$ strength distributions in $^{64}$Ni
compared with measured data \cite{Pop05}. The abscissa represents
energy in the daughter nucleus.} \label{fig:11}
\end{figure}

In Fig.~\ref{fig:11}, our model calculations for B(GT)$_{-}$
strength distributions in $^{64}$Ni have been compared with the
$^{64}$Ni$(^{3}He,t)$ charge-exchange reaction at 140 MeV/nucleon
\cite{Pop05}. Here we note that the pn-QRPA (C) model is in decent
agreement with measured data in low-energy region. Remaining QRPA
models calculate bulk of strength at high excitation energies in
daughter.

\begin{figure}
\begin{center}
\resizebox{1.25\textwidth}{!}{
\includegraphics{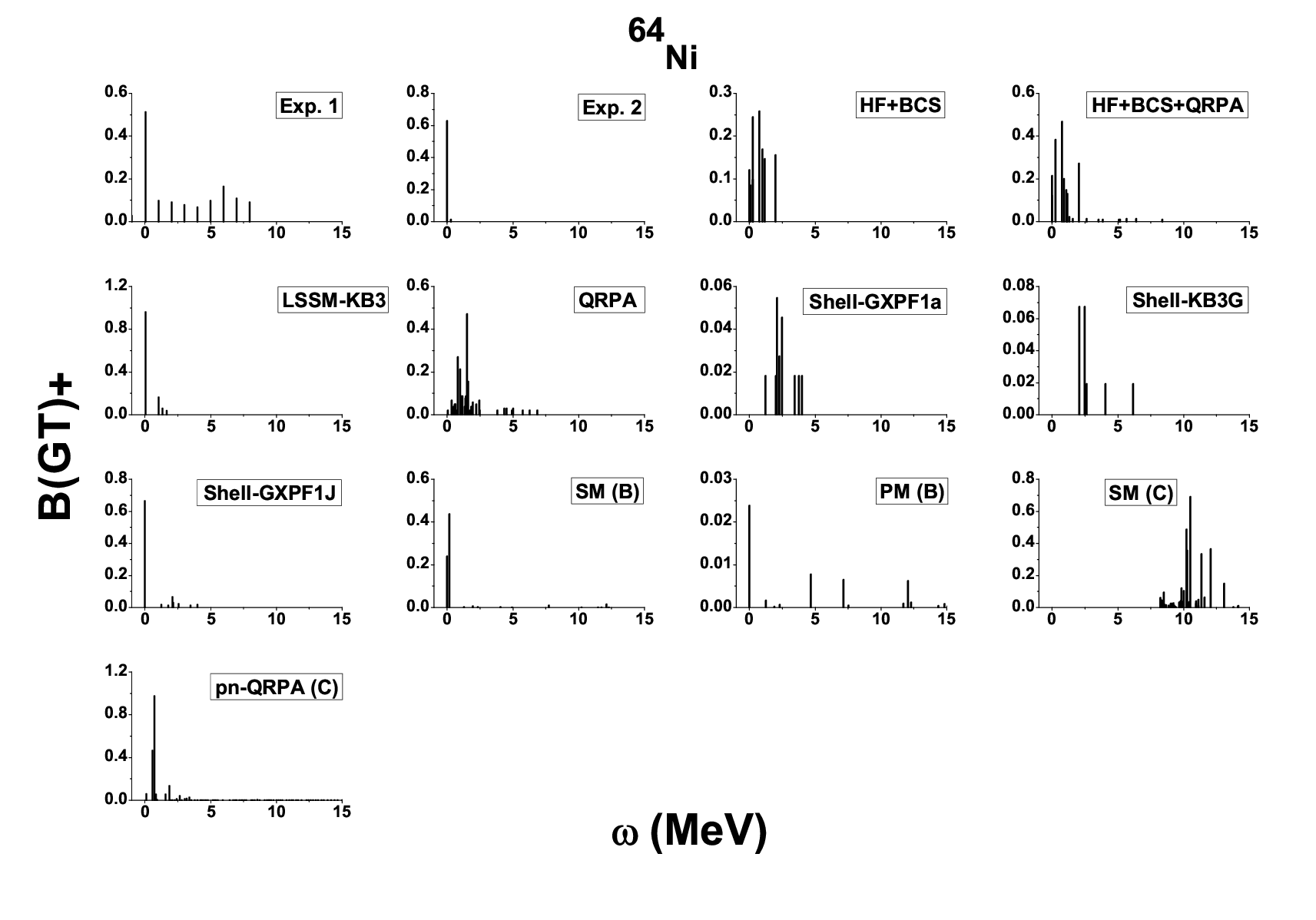}}
\end{center}
\caption{GT strength distributions in $^{64}$Ni along electron
capture direction. Exp. 1 shows measured distribution by
\cite{Will95} and Exp. 2 by \cite{Pop07}. HF+BCS and HF+BCS+QRPA
calculations were taken from \cite{Sarr03}. Large scale shell model
calculation using KB3 interaction was taken from \cite{Cau99}. QRPA
and shell model (employing KB3G and GXPF1a interactions)
calculations were taken from \cite{Col12}. Shell model calculation
using GXPF1J interaction was taken from \cite{Suz11}. Last four
panels depict calculated strength distributions using our four QRPA
models. The abscissa represents energy in the daughter nucleus.}
\label{fig:12}
\end{figure}

We present the $^{64}$Ni$(n,p)$  \cite{Will95} and
$^{64}$Ni$(d,^{2}He)$ data \cite{Pop07} in first two panels of
Fig.~\ref{fig:12}. The remaining panels show theoretical
calculations and are in same order as depicted in Fig.~\ref{fig:10}.
Here one notes that LSSM-KB3 and pn-QRPA (C) model calculations are
in good agreement with data of Exp. 1. The pn-QRPA (C) model
calculates low-lying centroid as compared to PM (B), PM (C) and SM
(C) models.

As mentioned earlier, with proceeding collapse the stellar density
increases by orders of magnitude and the resulting electron chemical
potential is significantly higher than the nuclear $Q$ value. Under
prevailing conditions the stellar rates are largely determined by
the  total strength and centroid of the GT$_\pm$ strength
distributions. In Tables.~\ref{tab:1}~-~\ref{tab:5} we present the
total B(GT) strength, along $\beta^{-}$ and $\beta^{+}$ directions,
and compare our calculated cumulative strength with previous
calculations and measurements wherever possible. It is noted that in
majority of the cases the total GT strength calculated by our
pn-QRPA(C) model is in very good agreement with measured data.

It can be seen from Tab.~\ref{tab:1} and Tab.~\ref{tab:2} that the
pn-QRPA(C) model calculates a bigger B(GT)$_{-}$ strength of 5.34
for  $^{56}$Ni as compared to the measured $^{56}$Ni$(p,n)$ data
\cite{Sas12} of 3.8 units. Shell model and RPA calculations estimate
even bigger values of total strength for  $^{56}$Ni. For $^{58}$Ni
and $^{60}$Ni, the pn-QRPA(C) calculated total B(GT)$_{-}$ strength
is in excellent agreement with measured data. Our model calculates
total strength of 6.89 units for the case of $^{58}$Ni to be
compared with the measured value of 7.4 \cite{Rap83}. In case of
$^{60}$Ni the calculated total strength value of 7.38 units is even
closer to the measured value of 7.2 units \cite{Rap83}. For
$^{64}$Ni the measured total strength is 1.9 units in the
$(^{3}He,t)$ experiment \cite{Pop05} up to 3.5 MeV excitation energy
in daughter. The pn-QRPA(C) model calculated strength up to 3.5 MeV
in $^{64}$Cu is 3.7 units (in Tab.~\ref{tab:2} the calculated value
of 10.5 is up to 20 MeV in daughter nucleus).

The calculated and measured total B(GT)$_{+}$ strength values are
shown in Tables.~\ref{tab:3}~-~\ref{tab:5}. Total strength for
$^{58}$Ni is shown in Tab.~\ref{tab:3} and Tab.~\ref{tab:4}. The
$^{58}$Ni$(t,^{3}He,)$$^{58}$Co data \cite{Col06} of 4.1 units is in
very good agreement with our pn-QRPA (C) value of 4.73 units.
Likewise for $^{60}$Ni the measured total strength of 3.11
\cite{Will95} compares excellent with the pn-QRPA(C) value of 3.06
units. For $^{62}$Ni ($^{64}$Ni) the $(n,p)$ data gives a total
strength of 2.5 (1.72) \cite{Will95}. This is to be compared with
the pn-QRPA(C) calculated value of 2.43 (1.86) units, respectively.
It is noted that the spectral distribution  \cite{Sar88} and shell
model calculations (using vector method code, FPC8 and FPVH
interactions) \cite{Will95,Bloom85} tend to overestimate the
calculated strength. Remaining shell model calculations are, in
general, very good agreement with measured data.

We finally present comparison of calculated and measured centroid
values in Tab.~\ref{tab:6} and Tab.~\ref{tab:7}. It is noted that
the pn-QRPA(C) model places centroid at relatively low energies in
daughter compared with other calculations. The centroid placement is
also decent when compared with experimental data.

\section{Summary and Conclusions}
\label{sec:3} GT transitions for $fp$-shell nuclei play a leading
role in the presupernova and supernova phases of massive stars. In
this work, we calculated GT strength distribution functions for
isotopes of nickel using microscopic QRPA methods (namely pn-QRPA,
Schematic Model and Pyatov Method).  Within this framework, we
studied the role of deformed basis in QRPA methods. We did our
calculations for 56$\leq$A$\leq$64.

Our calculations showed that the models with deformation of the
nucleus incorporated gave better results for total GT strength and
fulfillment of Ikeda sum rule as against those models performed in
spherical basis. The pn-QRPA (C) model was found the best amongst
the four models used. Not only were the model results in good
agreement with measured data but it also resulted in placement of
centroid at low excitation energies as compared to SM (C) model.
Tab.~\ref{tab:6} and Tab.~\ref{tab:7} confirm the fact reported in
\cite{Nab13} that the pn-QRPA (C) model places the centroid at much
lower energies in daughter nuclei as compared to other RPA
interactions. It was also concluded in \cite{Nab13} that GT centroid
placement by the pn-QRPA (C) model is, in general, in very good
agreement with the centroids of measured data. It was shown in the
current work that not only the low-lying measured GT distribution
was reproduced well by the pn-QRPA(C) model but also the calculated
total GT strength and centroid placement were in decent agreement
with measured data. Amongst all the RPA models the pn-QRPA(C) model
is the clear choice and should be used for a reliable estimate of
weak-interaction mediated rates in stellar environment.

Likewise, LSSM model results in very good agrement of calculated GT
strength distributions with measured data. However the pn-QRPA (C)
model has the additional feature that the model can be used for any
arbitrary heavy nucleus. This advantage comes handy when GT strength
distributions for hundreds of nuclei are required (including heavy
ones) for modeling of various astrophysical phenomena.

\vspace{0.5 in}\textbf{Acknowledgments:}   J.-U. Nabi wishes to
acknowledge the support provided by T\"{u}bitak (Turkey) under
Program No. 1059B211402772, Higher Education Commission Pakistan
through project number 5557/KPK/NRPU/R$\&$D/HEC/2016 and Pakistan
Science Foundation through project number PSF-TUBITAK/KP-GIKI (02).

\newpage
\newpage
\newpage
\newpage
\newpage
\newpage
\newpage
\newpage
\newpage
\newpage
\newpage
\newpage
\newpage
\newpage
\newpage
\newpage
\newpage
\newpage
\newpage
\newpage
\newpage
\newpage
\newpage
\newpage
\newpage
\newpage

\begin{table}
 \caption{Total B(GT$_{-})$ strength for nickel isotopes.}\label{tab:1}
\begin{tabular}{ll}
$^{56}$Ni  & $\sum B(GT)_{-}$\\
\hline
Exp. \cite{Sas12}&3.8  \\
HFB with RPA \cite{Bai13}&18.28  \\
RPA-GXPF1 \cite{Nab13}&7.58\\
RPA-FPD6 \cite{Nab13}&6.34\\
RPA-KB3G \cite{Nab13}&6.81\\
HF+RPA \cite{Ham93} &13.5\\
LSSM-KB \cite{Nak96}&11.4 \\
Shell-GXPF1 \cite{Suz11}&6.2 \\
Shell-GXPF1A \cite{Suz11}&6.2 \\
Shell-GXPF1J \cite{Suz11}&6.2 \\
Shell-KBF \cite{Suz11}&5.3 \\
Shell-KB3G \cite{Suz11}&5.4 \\
SM (B)&3.58\\
PM (B)&0.53\\
SM (C)&5.86\\
pn-QRPA (C)&5.34\\
\hline
$^{57}$Ni  & \\
\hline
LSSM-KB \cite{Nak96}&14.0 \\
SM (B)&1.56\\
PM (B)&0.57\\
SM (C)&6.2\\
pn-QRPA (C)&5.5\\
\hline
$^{58}$Ni  & \\
\hline
Exp.1 \cite{Rap83}&7.4 \\
Exp.2 \cite{Fuj02,Fuj07}&3.67 \\
RPA-GXPF1 \cite{Nab13}&9.7\\
RPA-FPD6 \cite{Nab13}&7.96\\
RPA-KB3G \cite{Nab13}&8.51\\
Shell-GXP1J \cite{Suz09}&8.0\\
LSSM-KB \cite{Nak96}&15.8\\
Spect.Dist.(configuration) \cite{Sar88}&19.33\\
Spect.Dist.(scalar) \cite{Sar88}&18.03\\
Shell Model \cite{Bloom85}&16.6\\
LSSM-KB3 \cite{Cau99}&7.7\\
pn-QRPA \cite{Sarr03}&16.46 \\
\end{tabular}
\end{table}
\clearpage
\begin{table}
 \caption{Same as Table~\ref{tab:1}.}\label{tab:2}
\begin{tabular}{ll}

$^{58}$Ni  & $\sum B(GT)_{-}$ \\
\hline
SM (B)&5.12\\
PM (B)&1.67\\
SM (C)&8.09\\
pn-QRPA (C)&6.89\\
\hline
$^{60}$Ni  & \\
\hline
Exp. \cite{Rap83}&7.2\\
Shell-GXP1J \cite{Suz09}&9.9\\
LSSM-KB3 \cite{Cau99}&10\\
pn-QRPA \cite{Sarr03}&19.71 \\
Spect.Dist.(configuration) \cite{Sar88}&23.51\\
Spect.Dist.(scalar) \cite{Sar88}&21.62\\
Shell Model \cite{Bloom85}&24.6\\
SM (B)&6.45\\
PM (B)&4.69\\
SM (C)&10.27\\
pn-QRPA (C)&7.38\\
\hline
$^{62}$Ni  & \\
\hline
pn-QRPA \cite{Sarr03}& 22.77\\
SM (B)&8.07\\
PM (B)&6.75\\
SM (C)&12.29\\
pn-QRPA (C)&8.9\\
\hline
$^{64}$Ni  & \\
\hline
Exp. \cite{Pop05}&1.9\\
pn-QRPA \cite{Sarr03}&27.24 \\
SM (B)&9.35\\
PM (B)&8.75\\
SM (C)&13.4\\
pn-QRPA (C)&10.5\\
\end{tabular}
\end{table}

\begin{table}
 \caption{Total B(GT$_{+})$ strength for nickel isotopes.}\label{tab:3}
\begin{tabular}{ll}

$^{56}$Ni  & $\sum B(GT)_{+}$ \\
\hline
RPA-GXPF1 \cite{Nab13}&7.58\\
RPA-FPD6 \cite{Nab13}&6.34\\
RPA-KB3G \cite{Nab13}&6.81\\
LSSM-KB \cite{Nak96}&11.4 \\
Shell-GXPF1\cite{Suz11}&6.2 \\
Shell-GXPF1A\cite{Suz11}&6.2 \\
Shell-GXPF1J\cite{Suz11}&6.2 \\
Shell-KBF\cite{Suz11}&5.3 \\
Shell-KB3G\cite{Suz11}&5.4 \\
SMMC \cite{Lan95}&9.86 \\
SM (B)&3.57\\
PM (B)&0.46\\
SM (C)&5.83\\
pn-QRPA (C)&5.34\\
\hline
$^{57}$Ni  & \\
\hline
LSSM-KB \cite{Nak96}&10.9 \\
SM (B)&1.29\\
PM (B)&0.33\\
SM (C)&5.56\\
pn-QRPA (C)&4.42\\
\hline
$^{58}$Ni  & \\
\hline
Exp.1 \cite{Kat94}&3.8 \\
Exp.2 \cite{Hag04}&2.04 \\
Exp.3 \cite{Col06}&4.1 \\
RPA-GXPF1 \cite{Nab13}&6.09\\
RPA-FPD6 \cite{Nab13}&4.36\\
RPA-KB3G \cite{Nab13}&4.91\\
Shell-GXP1J \cite{Suz09}&4.7\\
LSSM-KB\cite{Nak96}&9.5\\
LSSM-KB3\cite{Cau99}&4.4\\
SMMC \cite{Lan95}&6.72\\
Shell-FPC8 \cite{Will95}&12.17\\
Shell-FPVH \cite{Will95}&12.28\\
Spect.Dist.(configuration) \cite{Sar88}&13.33\\
Spect.Dist.(scalar) \cite{Sar88}&12.03\\
Shell Model \cite{Bloom85}&10.6\\
\end{tabular}
\end{table}
\begin{table}
 \caption{Same as Table~\ref{tab:3}.}\label{tab:4}
\begin{tabular}{ll}

$^{58}$Ni  & $\sum B(GT)_{+}$ \\
\hline
HF+BCS \cite{Sarr03}&6.19\\
pn-QRPA \cite{Sarr03}& 5.00\\
Shell-GXPF1\cite{Suz11}&4.7 \\
Shell-GXPF1A\cite{Suz11}&4.7 \\
Shell-GXPF1J\cite{Suz11}&4.7\\
Shell-KBF\cite{Suz11}&4.2 \\
Shell-KB3G\cite{Suz11}&4.0\\
GXPF1 \cite{Col06}&4.1\\
SM (B)&2.87\\
PM (B)&0.15\\
SM (C)&5.88\\
pn-QRPA (C)&4.73\\
\hline
$^{60}$Ni  & \\
\hline
Exp. \cite{Will95}&3.11\\
Shell-GXP1J \cite{Suz09}&3.4\\
Spect.Dist.(configuration) \cite{Sar88}&11.51\\
Spect.Dist.(scalar) \cite{Sar88}&9.62\\
Shell Model \cite{Bloom85}&12.6\\
SMMC \cite{Lan95} &5.18 \\
Shell-FPC8 \cite{Will95}&9.79\\
Shell-FPVH \cite{Will95}&9.86\\
LSSM-KB3 \cite{Cau99}&3.4\\
HF+BCS \cite{Sarr03}&4.97\\
pn-QRPA \cite{Sarr03}&3.72\\
Shell-GXPF1\cite{Suz11}&3.4\\
Shell-GXPF1A\cite{Suz11}&3.4 \\
Shell-GXPF1J\cite{Suz11}&3.4\\
Shell-KBF\cite{Suz11}&3.1 \\
Shell-KB3G\cite{Suz11}&2.8\\
SM (B)&2.17\\
PM (B)&0.27\\
SM (C)&5.68\\
pn-QRPA (C)&3.06\\
\end{tabular}
\end{table}
\clearpage
\begin{table}
 \caption{Same as Table~\ref{tab:3}.}\label{tab:5}
\begin{tabular}{ll}
$^{62}$Ni  & $\sum B(GT)_{+}$ \\
\hline
Exp. \cite{Will95}&2.5\\
Shell-FPC8 \cite{Will95}&7.67\\
Shell-FPVH \cite{Will95}&7.29\\
Shell-GXPF1\cite{Suz11}&2.0\\
Shell-GXPF1A\cite{Suz11}&1.9 \\
Shell-GXPF1J\cite{Suz11}&1.9\\
Shell-KBF\cite{Suz11}&2.0\\
Shell-KB3G\cite{Suz11}&2.0\\
LSSM-KB3 \cite{Cau99}&2.1\\
HF+BCS \cite{Sarr03}&3.40\\
pn-QRPA \cite{Sarr03}&2.36\\
SMMC \cite{Lan95} &3.43 \\
SM (B)&1.53\\
PM (B)&0.2\\
SM (C)&5.46\\
pn-QRPA (C)&2.43\\
\hline
$^{64}$Ni  & \\
\hline
Exp. 1 \cite{Will95}&1.72\\
Exp. 2 \cite{Pop07}&1.18\\
Shell-FPC8 \cite{Will95}&5.1\\
Shell-FPVH \cite{Will95}&4.57\\
Shell-GXPF1\cite{Suz11}&1.0\\
Shell-GXPF1A\cite{Suz11}&0.9 \\
Shell-GXPF1J\cite{Suz11}&0.9\\
Shell-KBF\cite{Suz11}&1.2\\
Shell-KB3G\cite{Suz11}&1.1\\
LSSM-KB3 \cite{Cau99}&1.3\\
HF+BCS \cite{Sarr03}&2.65\\
pn-QRPA \cite{Sarr03}&1.65\\
SMMC \cite{Lan95} &1.73 \\
SM (B)&0.73\\
PM (B)&0.06\\
SM (C)&4.76\\
pn-QRPA (C)&1.86\\
\end{tabular}
\end{table}
\begin{table}
 \caption{Centroid values of GT strength distributions in even-even
nickel isotopes.}\label{tab:6}
\begin{tabular}{lll}
$^{56}$Ni  & $\bar{E}_{m}$ [MeV]&$\bar{E}_{p}$ [MeV]\\
\hline
Exp. \cite{Sas12}&4.1 &- \\
RPA-GXPF1 \cite{Nab13}&11.54&11.54\\
RPA-FPD6 \cite{Nab13}&10.62&10.62\\
RPA-KB3G \cite{Nab13}&9.77&9.77\\
Shell-KB3G \cite{Lan98} &6.0&-\\
SMMC \cite{Dean98}&- &2.6\\
FFN \cite{Ful80}&- &3.78\\
Shell-GXPF1 \cite{Suz11}&- &5.2 \\
Shell-GXPF1A \cite{Suz11}&- &5.2 \\
Shell-GXPF1J \cite{Suz11}&-  &5.0\\
Shell-KBF \cite{Suz11}& -&4.4 \\
Shell-KB3G \cite{Suz11}&- &3.7 \\
SM (B)&3.86&4.36\\
PM (B)&5.65&7.58\\
SM (C)&3.66&19.76\\
pn-QRPA (C)&2.26&3.11\\
\hline
$^{58}$Ni  & & \\
\hline
Exp.1 \cite{Kat94}&- &4.0 \\
Exp.2 \cite{Col06}&- &4.4\\
Exp.3 \cite{Rap83}& 9.4&- \\
RPA-GXPF1 \cite{Nab13}&14.6&6.79\\
RPA-FPD6 \cite{Nab13}&13.94&6.15\\
RPA-KB3G \cite{Nab13}&13.97&5.02\\
SMMC \cite{Dean98}&- &2.0\\
LSSM-KB3 \cite{Cau99} &- &3.75\\
FFN \cite{Ful80}&6.5 &3.76\\
Shell-GXPF1 \cite{Suz11}&- & 4.2\\
Shell-GXPF1A \cite{Suz11}& -&4.3 \\
Shell-GXPF1J \cite{Suz11}& - &4.1\\
Shell-KBF \cite{Suz11}&- &3.7 \\
Shell-KB3G \cite{Suz11}& -&2.9 \\
SM (B)&7.01&3.54\\
PM (B)&4.15&4.93\\
SM (C)&9.27&17.82\\
pn-QRPA (C)&4.48&4.27\\
\hline
\end{tabular}
\end{table}
\begin{table}
 \caption{Same as Table~\ref{tab:6}.}\label{tab:7}
\begin{tabular}{lll}
$^{60}$Ni  & $\bar{E}_{m}$ [MeV] &$\bar{E}_{p}$ [MeV]\\
\hline

Exp. \cite{Rap83}&9.0&-\\
SMMC \cite{Dean98}& -&1.0\\
FFN \cite{Ful80}&4.9&2.0\\
LSSM-KB3 \cite{Cau99}&-&2.88\\
Shell-GXPF1 \cite{Suz11}&- &3.0 \\
Shell-GXPF1A \cite{Suz11}&- &3.1 \\
Shell-GXPF1J \cite{Suz11}&-  &2.8\\
Shell-KBF \cite{Suz11}&- &2.7 \\
Shell-KB3G \cite{Suz11}&- &2.4 \\
SM (B)&6.97&2.39\\
PM (B)&6.54&4.8\\
SM (C)&13.87&16.67\\
pn-QRPA (C)&4.27&3.18\\
\hline
$^{62}$Ni  & & \\
\hline
LSSM-KB3 \cite{Cau99}& -&1.78\\
FFN \cite{Ful80}&-&2.0\\
Shell-GXPF1 \cite{Suz11}&- & 1.8\\
Shell-GXPF1A \cite{Suz11}&- &2.0 \\
Shell-GXPF1J \cite{Suz11}&-  &1.8 \\
Shell-KBF \cite{Suz11}&- & 1.7 \\
Shell-KB3G \cite{Suz11}&- & 1.5 \\
SM (B)&8.27&1.76\\
PM (B)&6.28&5.77\\
SM (C)&15.09&13.39\\
pn-QRPA (C)&4.55&2.28\\
\hline
$^{64}$Ni  & & \\
\hline
LSSM-KB3 \cite{Cau99}& -&0.5\\
FFN \cite{Ful80}&-&2.0\\
Shell-GXPF1 \cite{Suz11}&- & 0.8 \\
Shell-GXPF1A \cite{Suz11}&- &0.9 \\
Shell-GXPF1J \cite{Suz11}&-  & 0.8 \\
Shell-KBF \cite{Suz11}&- &0.5 \\
Shell-KB3G \cite{Suz11}&- &0.5 \\
SM (B)&9.41&1.29\\
PM (B)&6.63&6.68\\
SM (C)&19.66&13.68\\
pn-QRPA (C)&4.93&1.0\\
\end{tabular}
\end{table}

\end{document}